\documentclass[runningheads]{llncs}
\usepackage[T1]{fontenc}
\usepackage{amsmath,amssymb,amsfonts}
\usepackage{algorithmic}
\usepackage{graphicx}
\usepackage{multirow}
\usepackage{textcomp}
\usepackage{xcolor}
\usepackage{xspace}
\usepackage{listings}
\usepackage{hyperref}
\usepackage{tikz}
\usepackage{mathpartir}
\usetikzlibrary{arrows.meta, calc, positioning, shadows}

\definecolor{gcsblue}{RGB}{230, 240, 255}
\definecolor{uavblue}{RGB}{230, 240, 255}
\definecolor{msgblue}{RGB}{0, 102, 204}
\definecolor{msggreen}{RGB}{34, 139, 34}

\definecolor{protocolBlue}{RGB}{0, 102, 204}
\definecolor{protocolGreen}{RGB}{0, 153, 76}

\lstdefinelanguage{Fstar}{
  morekeywords={module, let, type, val, assume, begin, end, function, forall, fun, inline_for_extraction, noeq, and, match, with, if, then, else, Tot, Lemma, requires, ensures, decreases},
  sensitive=true,
  morecomment=[l]{//},
  morecomment=[s]{(*}{*)},
  morestring=[b]",
  literate={-->}{{$\longrightarrow$}}3 {->}{{$\to$}}2
}

\lstdefinelanguage{C}{
  morekeywords={switch, case, break, default, if, else, return, true, false, bool, typedef, enum, struct, int, void, const},
  sensitive=true,
  morecomment=[l]{//},
  morecomment=[s]{/*}{*/},
  morestring=[b]",
}

\begin{document}
\title{From High-Level Types to Low-Level Monitors: Synthesizing Verified Runtime Checkers for MAVLink}
\titlerunning{From High-Level Types to Low-Level Monitors}

\author{ 
Arthur Amorim\inst{1}\orcidID{0009-0003-7712-5055} \and 
Paul Gazzillo\inst{1}\orcidID{0000-0003-1425-8873} \and \\
Max Taylor\inst{2} \orcidID{0009-0005-7873-9694} \and 
Lance Joneckis\inst{3}\orcidID{0009-0002-0284-4787}
}

\authorrunning{A. Amorim et al.}


\institute{ 
University of Central Florida, Florida, USA\\
\email{\{arthur.amorim, paul.gazzillo\}@ucf.edu}
\and 
Boise State University, Boise, USA \\
\email{maxtaylor@boisestate.edu}
\and
Idaho National Laboratory, Idaho Falls, USA\\
\email{lance.joneckis@inl.gov}
}

\maketitle              % typeset the header of the contribution
\begin{abstract}
Standard communication protocols for Unmanned Aerial Vehicles (UAVs), such as MAVLink, lack the capability to enforce the contextual validity of message sequences.
Autopilots therefore remain vulnerable to stealthy attacks, where syntactically correct but semantically ill-timed commands induce unsafe states without triggering physical anomaly detectors.
Prior work (DATUM) demonstrated that \emph{global} Refined Multiparty Session Types (RMPSTs) are an effective specification language for centralized MAVLink protocol enforcement, but suffered from two engineering failures: manual proof terms interleaved with protocol definitions, and an OCaml extraction backend whose managed runtime is incompatible with resource-constrained UAV hardware.
We present \textsc{Platum}, a framework that addresses both failures with a minimal DSL requiring only the five semantic components of a global session type (sender, receiver, label, payload variable, refinement predicate), whose structural well-formedness conditions are confirmed via reflective decision procedures in Meta-F*.
Confirmed specifications are compiled directly into flat, allocation-free C Finite State Machines (FSMs), deployed as centralized proxy monitors at the GCS/UAV communication boundary.
Our evaluation demonstrates a $4\times$ reduction in total monitor latency and lower memory overhead compared to DATUM, measured via ArduPilot SITL simulation.

\end{abstract}

 \section{Introduction}

% Idea: UAVs are critical infrastructure relying on MAVLink, which lacks inherent security for conversation semantics.
Unmanned Aerial Vehicles (UAVs) have evolved from niche operational devices into essential tools, solving complex logistical and surveillance problems in sectors ranging from precision agriculture to national defense.
As these systems scale in autonomy and ubiquity, they increasingly rely on standardized communication backbones like MAVLink to coordinate with Ground Control Stations (GCS) and other aircraft~\cite{mavlinkdevelopmentteamMAVLinkMicroAir2024}.
However, while MAVLink excels at bandwidth efficiency, it was not designed with security as a primary constraint.
It effectively defines the syntax of the messages that can be sent, but fails to prescribe their semantics.

% Idea: The primary threat is the "stealthy attack" where insiders use valid syntax to enact malicious logic.
As a result, the security of UAV communication protocols has attracted significant research attention, with investigations revealing fundamental vulnerabilities in MAVLink's design and implementation~\cite{allouchMAVSecSecuringMAVLink2019,hamzaMAVLinkProtocolSurvey2024,kwonEmpiricalAnalysisMAVLink2018,taylorAvisInsituModel2021,kimRVFuzzerFindingInput2019,kimPGFUZZPolicyGuidedFuzzing2021}.
The primary threat addressed in this work is the \emph{stealthy attack}, in which a malicious actor (such as a disgruntled employee operating the GCS) sends messages that are syntactically correct, yet cause catastrophic damage to the system state~\cite{amorimEnforcingMAVLinkSafety2025,amorimUAVResilienceStealthy2025}.

% Idea: Prior work (formal and physics-based) exists but leaves a gap regarding logic attacks that are physically valid.
As a response to these stealthy attacks, researchers have explored formal verification and anomaly detection angles.
Physics-based anomaly detection methods, such as the R2U2 framework~\cite{R2U2MonitoringDiagnosis}, compare observed sensor data against physical system models.
While these methods are effective against disruptions to flight stability, they cannot detect logic-based attacks where the physical behavior remains nominal.

% Idea: Physics-based monitoring fails on logic errors, such as the MAVLink mission bug, necessitating protocol-level monitoring.
The limitations of physics-based monitoring are best illustrated by a previous vulnerability in the MAVLink mission sub-protocol~\cite{amorimEnforcingMAVLinkSafety2025,amorimUAVResilienceStealthy2025}:
\begin{center}
\begin{tikzpicture}[
    node distance=6cm,
    font=\sffamily\small,
    actor/.style={rectangle, draw, fill=blue!10, minimum width=2.5cm, minimum height=0.8cm, rounded corners, thick},
    lifeline/.style={dashed, thick, gray},
    message/.style={-Stealth, thick},
    loopbox/.style={draw, fill=gray!5, fill opacity=0.3, inner sep=5pt, dashed}
]
    % --- Actors --
    \node[actor] (GCS) {GCS};
    \node[actor, right=of GCS] (UAV) { UAV};

    % --- Defining Vertical Coordinates (Timesteps) ---
    \coordinate (ts1) at ($(GCS.south)-(0, 0.5)$);
    \coordinate (ts2) at ($(GCS.south)-(0, 1.5)$);
    \coordinate (ts3) at ($(GCS.south)-(0, 2.3)$);
    \coordinate (ts4) at ($(GCS.south)-(0, 3.1)$);
    \coordinate (ts5) at ($(GCS.south)-(0, 4.3)$);
    \coordinate (tsEnd) at ($(GCS.south)-(0, 5.2)$);

    % --- Lifelines ---
    \draw[lifeline] (GCS.south) -- (GCS.south |- tsEnd);
    \draw[lifeline] (UAV.south) -- (UAV.south |- tsEnd);

    % --- Protocol Steps ---
    % 1. MISSION_COUNT
    \draw[message, msgblue] (GCS.south |- ts1) -- node[above] {\texttt{MISSION\_COUNT (N)}} (UAV.south |- ts1);

    % 2. Loop Logic
    \draw[thick] ($(GCS.south |- ts2) + (-0.5, 0.3)$) rectangle ($(UAV.south |- ts4) + (0.5, -0.2)$);
    \node[anchor=north west, font=\sffamily\bfseries\scriptsize] 
        at ($(GCS.south |- ts2) + (+0.2, 0.3)$) {LOOP N times};

    % Loop Content
    \draw[message, msggreen] (UAV.south |- ts3) -- node[above] {\texttt{MISSION\_REQUEST\_INT (x)}} (GCS.south |- ts3);
    \draw[message, msggreen] (GCS.south |- ts4) -- node[above] {\texttt{MISSION\_ITEM\_INT (x)}} (UAV.south |- ts4);

    % 3. Final ACK
    \draw[message, msgblue] (UAV.south |- ts5) -- node[above] {\texttt{MISSION\_ACK (if x$=$curr then SUCCESS else ERROR)}} (GCS.south |- ts5);
\end{tikzpicture}
\end{center}
The mission upload sub-protocol dictates that a Ground Control Station (GCS) must initiate a mission upload by declaring a \lstinline{MISSION_COUNT} of $N$ items, implying the UAV should expect exactly $N$ subsequent items.
The items must be sent and requested in order, and the GCS must only send a \lstinline{MISSION_ACK (Success)} message if all items were correctly uploaded.
However, real-world implementations have failed to enforce this logical constraint. 
An issue report for the ArduPilot flight control software revealed that a UAV could accept fewer items than declared due to buffer mismanagement from previous flights~\cite{ardupilotMissionPlannerIssue12482024}.
This logical vulnerability allowed the UAV to execute a hybrid flight plan mixing old and new waypoints, compromising the mission without violating physical laws.
Preventing such vulnerabilities requires rigorous runtime monitoring of the protocol logic itself.

% Idea: DATUM successfully used Refined Multiparty Session Types to solve these logic errors.
To address these logical vulnerabilities, the DATUM framework~\cite{amorimEnforcingMAVLinkSafety2025} demonstrated the promise of using \emph{global} Refined Multiparty Session Types (RMPSTs) as a specification language for centralized proxy-based enforcement.
DATUM formalizes MAVLink in the F* theorem prover~\cite{swamyDependentTypesMultimonadic2016}, enabling the generation of centralized runtime monitors that intercept logical vulnerabilities before they result in undefined behavior.
Consequently, DATUM does not alter the MAVLink protocol or rely on the MAVLink development team for bug fixes; it acts as a transparent proxy, allowing operators to enforce arbitrarily strict protocol contracts at the GCS/UAV link.

% Idea: DATUM has critical limitations: arduous manual proofs (intrinsic verification) and high memory overhead (OCaml extraction).
Despite its theoretical success, the DATUM framework suffers from two critical limitations that hinder practical deployment.
First, the specification language conflates protocol definition with well-formedness proof. Because DATUM enforces structural invariants \emph{intrinsically}, through the F* type system during term construction, users must supply proof terms, accumulator arguments, and label history threads as part of the session type itself. The five semantic components of a session type (sender, receiver, label, payload variable, and refinement predicate) are obscured beneath a significant burden of proof scaffolding, producing prohibitively long compilation times and a high barrier to entry for protocol engineers.
Second, DATUM relied on OCaml extraction for its centralized runtime monitors, as the standard extraction language for most theorem provers.
The OCaml runtime environment introduces a Resident Set Size (RSS) and performance overhead that is not optimal for the resource-constrained micro-controllers typically found in UAVs.

% Idea: Platum solves this by moving to Extrinsic Verification (Meta-F*) and Synthesizing C.
In this paper, we present \textsc{Platum} (Performant Lightweight Assured Typed Universal Messaging), an enhanced framework that addresses both the language design and performance gaps of DATUM.
The key architectural decision in \textsc{Platum} is the separation of specification from well-formedness checking. The user supplies only the five semantic components of a global session type. A suite of reflective decision procedures, implemented via Meta-F*'s~\cite{martinezMetaFProofAutomation2019} AST inspection capabilities, then confirms structural invariants -- label uniqueness, guarded recursion, global progress, and transition fidelity -- algorithmically after the session type is defined. No proof terms, accumulator arguments, or label history management appear in the user-facing DSL.

Furthermore, \textsc{Platum} introduces a direct synthesis pipeline that translates structurally confirmed session type ASTs into standalone C source code, deployed as a centralized proxy monitor at the GCS/UAV communication boundary.
Unlike general-purpose extraction chains that rely on intermediate languages like Low* to prove memory safety for complex heap manipulations, \textsc{Platum} targets a specific subset of C: flat Finite State Machines (FSMs).
By mapping refined global session types directly to C switch statements and control-flow checks, we generate native centralized runtime monitors that eliminate the need for an OCaml runtime or garbage collection, enabling deployment on the resource-constrained ARM micro-controllers found in UAV flight controllers.

% Idea: Summary of contributions (DSL, Synthesis pipeline, Evaluation).
To summarize our contributions:
\begin{itemize}
    \item \textbf{A minimal global RMPST DSL with algorithmic well-formedness checking (Section~\ref{sec:methodology}).} The \textsc{Platum} DSL requires only the five semantic components of a session type: sender, receiver, label, payload variable, and refinement predicate. A suite of reflective decision procedures in Meta-F* then confirms structural invariant on the AST, with no proof terms required from the user.
    \item \textbf{A direct C FSM synthesis pipeline for centralized enforcement (Section~\ref{sec:synthesis}).} Structurally confirmed global session types are translated into flat, allocation-free C Finite State Machines and deployed as proxy monitors at the GCS/UAV boundary, eliminating the OCaml runtime and its garbage collection overhead.
    \item \textbf{A quantitative evaluation against DATUM (Section~\ref{sec:evaluation}),} demonstrating a $4\times$ reduction in total monitor latency and lower RSS overhead via ArduPilot SITL experiments.
\end{itemize}

\section{Overview}
\begin{figure}[t]
    \centering
    \includegraphics[width=0.8\linewidth]{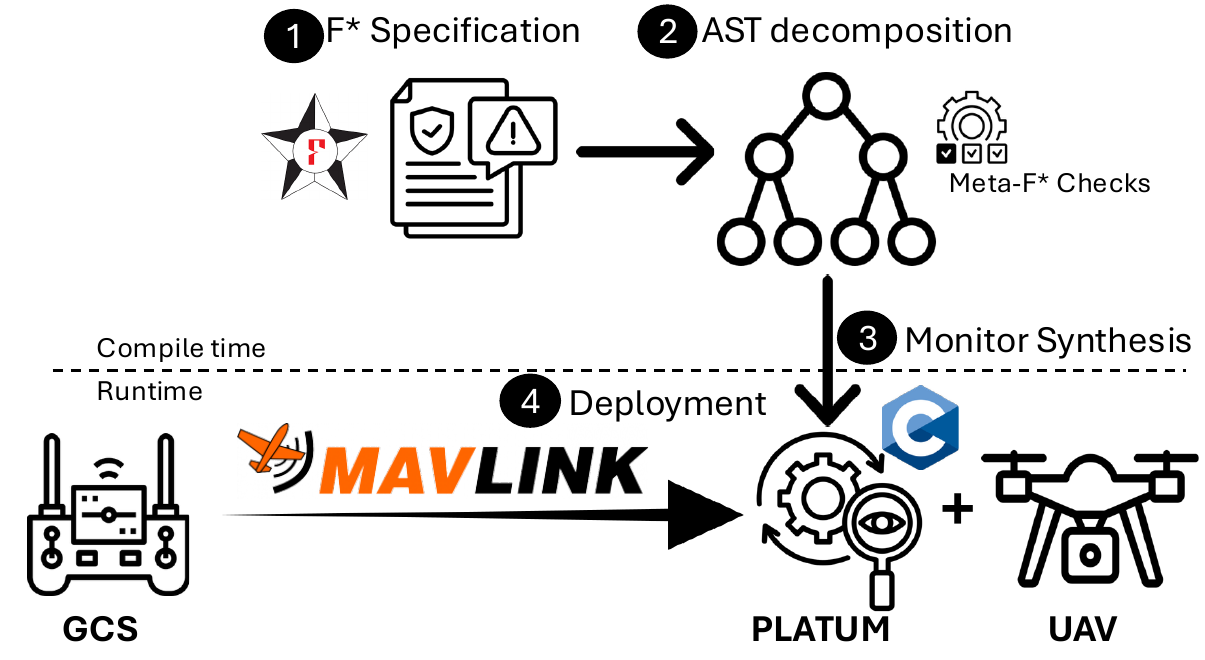}
    \caption{The \textsc{Platum} Workflow. The compile-time pipeline (above the dotted line) proceeds from F* specification, through AST decomposition and Meta-F* well-formedness checking, to C monitor synthesis. The runtime deployment (below the dotted line) shows the synthesized monitor operating as a centralized proxy: the GCS transmits MAVLink traffic to \textsc{Platum}, which enforces the global session type at the network boundary before relaying conformant messages to the UAV autopilot.}
    \label{fig:workflow}
\end{figure}

\subsection{A Birds-Eye View of \textsc{Platum}}

\textsc{Platum} addresses DATUM's limitations through a design centered on a single principle: the user supplies only the five semantic components of a global session type, sender, receiver, label, payload variable, and refinement predicate, and the framework handles the rest.
As shown in Figure~\ref{fig:workflow}, the workflow separates compile-time analysis from runtime enforcement and proceeds in four stages:

\begin{enumerate}
    \item \textbf{Specification:} The user defines the MAVLink protocol using our F* DSL. This specification includes the message structures (derived from standard MAVLink XML definitions via a Python transpiler) and the protocol logic expressed as a global session type. No proof terms, accumulator arguments, or label history threads appear in user-facing code.
    \item \textbf{Reflection \& Well-Formedness Checking:} Meta-F* inspects the AST of the session type after it is defined. A suite of reflective decision procedures, implemented as total boolean functions, confirms structural invariants: label uniqueness, guarded recursion, global progress, and transition fidelity. These checks require no input from the user beyond the session type itself.
    \item \textbf{Monitor Synthesis:} Once the structural invariants are confirmed, the AST is translated directly into a flat C Finite State Machine (FSM). Refinement predicates are compiled into C conditional guards; recursive binders are compiled into context variable updates. No dynamic memory allocation is introduced.
    \item \textbf{Deployment:} The synthesized monitor is compiled as a shared library and deployed as a centralized proxy at the GCS/UAV communication boundary, without modifying flight controller firmware. A single monitor instance enforces the global session type for all protocol traffic at the network boundary, with no per-participant instrumentation required.
\end{enumerate}

Sections~\ref{sec:methodology} and~\ref{sec:synthesis} detail the compile-time pipeline. Section~\ref{sec:evaluation} evaluates the runtime deployment against DATUM.

\subsection{Refined Multiparty Session Types (RMPSTs)}

\textsc{Platum} uses global Refined Multiparty Session Types (RMPSTs)~\cite{zhouStaticallyVerifiedRefinements2020,vassorRefinementsMultipartyMessagePassing2024} as a specification language: a global type describes the entire multiparty interaction as a single artifact, without specifying how individual participants are implemented.
This global view enables centralized enforcement, the monitor checks conformance at the network boundary without access to participant internals.
RMPSTs extend standard session types with predicate logic, allowing constraints to be placed directly on payload values, which is essential for MAVLink, where safety depends on the runtime values carried within messages rather than their structure alone.

We can formalize the MAVLink Mission Protocol using three core RMPST constructs:

\begin{itemize}
    \item \textbf{Global Interaction:} Sender $A$ transmits a refined message to $B$. The GCS opens the protocol by sending a count $N$, refined to lie within buffer limits:
    \[
      \texttt{GCS} \to \texttt{UAV}: \texttt{MISSION\_COUNT}(n: \mathbb{Z}\{ 1 \le n < \texttt{LIMIT} \})
    \]

    \item \textbf{Recursion ($\mu$):} The $\mu$ operator defines a loop state parameterized by index variables. Below, $\mu.\textbf{T}$ tracks the current item index:
    \[
       \mu.\textbf{T}(\textit{curr}: \mathbb{Z} \{ 0 \leq \textit{curr} < n\})
    \]

    \item \textbf{Branching and Choice:} The $\oplus$ operator represents internal choice; the sender selects a branch whose refinement constraint is satisfied. In the mission protocol, the UAV either requests the next item or acknowledges completion.
\end{itemize}

The complete, value-dependent formalization of the mission protocol appears in Figure~\ref{fig:running_example}; \textsc{Platum} uses this specification to generate a centralized monitor that confirms adherence at runtime.

\begin{figure}[t]
\centering
\small
\[
\begin{aligned}
   &\textbf{GCS} \to \textbf{UAV}\textbf{:}\ \texttt{MISSION\_COUNT}(n: \texttt{mission\_count}\ \{1 \leq \texttt{count}(n) < \texttt{LIMIT}\})\\
   &\mu.\textbf{T}(\textit{curr}: \mathbb{Z} \{ 0 \leq \textit{curr} \land \textit{curr} < \texttt{n}\})\ (\textit{curr} = 0) \\
   & \textbf{UAV} \to \textbf{GCS}\textbf{:} \\
      & \quad \texttt{MISSION\_REQUEST\_INT}(req: \texttt{mission\_request\_int}\ \{\texttt{req} = \textit{curr} \land \textit{curr} < \texttt{n}\}). \\
      &\quad \quad  \quad  \texttt{MISSION\_ITEM\_INT}(item: \texttt{mission\_item\_int}\ \{\texttt{item} = \textit{curr} \land \textit{curr} < \texttt{n}\}). \\
      &  \quad   \quad   \quad   \quad  \textbf{T}(\textit{curr}+1) \\
      & \quad  \oplus\ \texttt{MISSION\_ACK}(ack: \texttt{mission\_ack}\ \{\texttt{type}(ack) = \texttt{ERROR} \lor \textit{curr} = \texttt{n}\}).\ \textbf{End}
\end{aligned}
\]
\normalsize
\caption{RMPST Formalization of the MAVLink Mission Protocol}
\label{fig:running_example}
\end{figure}

\subsection{The DATUM Approach and its Limitations}
 
DATUM demonstrated that global RMPSTs could effectively model MAVLink's complex logic, but its intrinsic verification strategy severely hampered both usability and performance.
 
\subsubsection{The Burden of Intrinsic Verification}
 
The primary distinction between DATUM and \textsc{Platum} lies in how they enforce well-formedness.
A global session type has five semantic components: \textit{Sender}, \textit{Receiver}, \textit{Label}, \textit{Payload Variable}, and \textit{Refinement}.
Ideally, a DSL requires only these five components to construct the corresponding session type.
However, under intrinsic verification, the type system must be supplied with enough context to discharge well-formedness obligations at construction time, forcing the user to provide extra arguments that are irrelevant to session type theory but required for the check to go through.
 
\small
\begin{lstlisting}[caption={The heavy syntactic surface of DATUM}, label={lst:datum_mess}]
| Choice :  (#nparticipants : nat) ->
            (#rec_contexts : list (session_type u)) ->
            (from : nat) -> (to : nat) ->
            (l : (list (label * (dtuple4 ...
               (fun l k st -> proto_val st -> global_session))
               {no_duplicate_labels l}) ->
            global_session u ...
\end{lstlisting}
\normalsize

DATUM uses such intrinsic verification to confirm a property like ``no duplicate labels,''; the \texttt{Choice} constructor requires the user to pass accumulated label history and session continuations as explicit arguments (Listing~\ref{lst:datum_mess}).
The type signature exposes an accumulator ($l$), a recursion context stack ($rec\_contexts$), and a kind index ($k$), none of which are semantic components of the session type; they are proof scaffolding.
The refinement \texttt{\{no\_duplicate\_labels l\}} forces the SMT solver to discharge a well-formedness obligation over the entire label history on every branch addition, coupling every specification edit to a proof re-check.
\textsc{Platum} takes an extrinsic approach, where the DSL accepts only the five semantic inputs, and initial type checking performs only basic syntax validation.
A suite of reflective decision procedures in Meta-F* then traverses the AST \emph{after} the session type is defined, confirming structural conditions algorithmically on the term representation, with no accumulator arguments, no label history threads, and no proof scaffolding visible to the user.

\subsubsection{The Deployment Gap: OCaml vs. Embedded Systems}
 
The second limitation of DATUM lies in its code generation backend.
DATUM extracts to OCaml, whose runtime relies on a Garbage Collector (GC) and significant runtime type information, resulting in a high RSS memory footprint and non-deterministic performance.
For the embedded micro-controllers found on UAVs (such as the STM32 processors running PX4), this overhead could render the centralized monitors undeployable on critical system components.
Section~\ref{sec:synthesis} details how \textsc{Platum}'s direct C synthesis pipeline eliminates this overhead entirely.

%%% Local Variables:
%%% mode: LaTeX
%%% TeX-master: t
%%% End:
\section{Methodology}
\label{sec:methodology}

\subsection{Formalization of Protocol Logic}

We formalize Refined Multiparty Session Types (RMPSTs) using a minimal deep embedding in F* that focuses strictly on the semantic necessities: refined messages, branching, and recursion.
This design choice shifts the burden of well-formedness checking from the user, who no longer needs to construct complex proof terms manually, to the framework's algorithmic reflection capabilities.
Our formalization isolates the protocol logic into five components: the sender, the receiver, the message label, the payload variable, and the refinement predicate.

In the global session type formalism, the arrow notation $p_i \to p_j$ describes an interaction from above: participant $p_i$ sends and participant $p_j$ receives, as observed from a third-party view of the protocol.
This global perspective is fundamental to \textsc{Platum}'s enforcement model: the synthesized monitor occupies exactly this third-party position at the network boundary.

We define the syntax of our global session types $G$ inductively.
Let $\mathcal{P}$, $\mathcal{L}$, and $\mathcal{T}$ be the sets of participants, message labels, and refined types, respectively.
\begin{alignat*}{2}
G &::= \text{end} && \text{(Termination)} \\
  &\quad \mid p_i \to p_j : \{ l_k(x_k : \tau_k) \to G_k \}_{k \in K} && \text{(Choice)} \\
  &\quad \mid \mu X(v : \tau). G && \text{(Recursion)} \\
  &\quad \mid \texttt{recur } X(e) && \text{(Loop)}
\end{alignat*}

Each branch in a \textit{Choice} carries a label $l_k$ and a payload $x_k$ of type $\tau_k$, which may be a refined type (e.g., $x:\mathbb{Z}\{x > 0\}$), allowing protocol flow to depend on runtime values.
The \textit{Recursion} operator $\mu$ binds a variable $X$ and an initialization argument $v$; \textit{Loop} jumps back to the header with a new argument $e$.
Refinement expressions $e$ range over values, variables, arithmetic ($+, -, *, /$), relational ($=, <, >, \neq$), and boolean ($\neg, \wedge, \vee$) operators, covering all constraint forms that arise in MAVLink payload validation.

\subsection{Integrating MAVLink Specifications}

Since MAVLink protocols are defined in XML files that lack formal semantics, we must bridge the gap between these loose definitions and our type system.
We employ a Python-based transpiler that parses standard MAVLink XML definitions and generates a corresponding F* module of strongly typed tuples.
This constitutes part of the specification infrastructure: the transpiler handles the translation from XML field lists so that the engineer can supply the five semantic inputs per interaction step without manually re-encoding field layouts.
For example, the standard \texttt{MISSION\_COUNT} message is transformed from a flat XML field list into a refined tuple, allowing us to enforce constraints like buffer limits directly in the type signature:
$$
\texttt{mission\_count} \triangleq (\texttt{u8} \times \texttt{u8} \times \texttt{u16} \times \texttt{mav\_type} \times \texttt{u32})
$$

In Listing~\ref{lst:mavlink_spec}, we define the MAVLink Mission Upload Protocol using the \textsc{Platum} DSL.
Participant $0$ is the GCS and participant $1$ is the UAV.
Continuations are modeled as F* functions rather than inductive constructors: a session type consumes an argument to produce the subsequent state.
This is also the key technical enabler for the well-formedness checks in Section~\ref{subsec:wellformedness}, since Meta-F* can apply a continuation to a fresh symbolic variable and inspect the resulting term, allowing the decision procedures to traverse the protocol through recursive binders at specification time.

\begin{figure}[t]
    \centering
\begin{lstlisting}[language=Fstar, basicstyle=\footnotesize\ttfamily, caption={The MAVLink Mission Upload Protocol formalized in \textsc{Platum} (variable names are abbreviated due to spacing constraints).}, label={lst:mavlink_spec}]
let mavlink_mission_upload : session_type =
  (0 --> 1) [
    Option "MISSION_COUNT" (fun ((ts, tc, cnt, mt, oid) : mission_count {
      cnt >= 1 && cnt < mission_item_limit
    }) ->
      Mu 0 (fun (curr : int {0 <= curr && curr <= cnt}) ->
        (1 --> 0) [
          Option "MISSION_REQUEST_INT" (fun ((req_ts, ..., req_mt) : mission_request_int {
            req_seq = curr && curr < cnt
          }) ->
            (0 --> 1) [
              Option "MISSION_ITEM_INT" (fun (...) -> Recur 0 (curr + 1));
            ]);
          Option "MISSION_ACK" (fun (...) -> End)
        ]))]
\end{lstlisting}
\end{figure}

Standard avionics definitions like MAVLink's XML define the \textit{shape} of data but rarely the \textit{rules} of engagement.
Implicit constraints such as "this field must match the previous message's sequence number" are buried in prose documentation.
\textsc{Platum} makes these rules explicit via refinements, ensuring the formal model is a semantic enhancement that captures protocol intent in a machine-checkable format.

\subsection{Algorithmic Well-Formedness Checking via Meta-F*}
\label{subsec:wellformedness}
 
Checking structural invariants on session types that evolve as functions is syntactically burdensome if attempted intrinsically, as illustrated by DATUM's \texttt{Choice} constructor (Listing~\ref{lst:datum_mess}).
\textsc{Platum} circumvents this by adopting an extrinsic well-formedness checking strategy enabled by Meta-F*: we inspect the AST of the session type \emph{after} it has been defined, treating the type definition as data that can be algorithmically analyzed.
 
Concretely, the session type AST is first materialized into an intermediate representation graph: a finite directed structure in which nodes correspond to protocol states and edges correspond to labeled, refined transitions.
The four checks below operate on this graph; Section~\ref{subsec:extraction} details its construction.
Each check is implemented as a total recursive function over this graph, annotated with the \texttt{Tot} effect to guarantee termination.
These are boolean decision procedures: they return \texttt{true} if and only if the invariant holds.
The synthesis pipeline requires all four procedures to return \texttt{true} before any C code is emitted, enforcing that only structurally sound session types can be used for centralized monitoring.
 
Together, these four structural conditions are necessary for the synthesis pipeline to produce a well-defined C FSM: label uniqueness eliminates non-determinism at branch points, guarded recursion eliminates unproductive loops, global progress eliminates stuck states, and session fidelity eliminates dangling transitions.
 
\bigskip
\noindent\textbf{Check 1: Label Uniqueness.}\quad
Ambiguity in protocol design leads to non-deterministic behavior, which is unacceptable in safety-critical systems.
The decision procedure $\mathcal{A}_{unique}$ traverses the intermediate representation graph, accumulating a set of seen labels $\Sigma$, and returns \texttt{false} immediately upon finding a duplicate:
$$
\forall G.\ \mathcal{A}_{unique}(G, \emptyset) = \texttt{true} \implies \forall l_i, l_j \in \text{branches}(G).\ i \neq j \implies L_i \cap L_j = \emptyset
$$
Listing~\ref{lst:unique_labels} shows the concrete F* implementation.
 
\begin{lstlisting}[language=Fstar, basicstyle=\footnotesize\ttfamily, caption={The label uniqueness decision procedure. The \texttt{Tot} annotation guarantees termination; \texttt{seen} corresponds to the accumulator set $\Sigma$.}, label={lst:unique_labels}]
let rec has_duplicates (l: list label) (seen: list label) : Tot bool =
  match l with
  | [] -> false
  | hd :: tl ->
      if L.mem hd seen && hd <> "RECUR" then true
      else has_duplicates tl (hd :: seen)
 
let check_unique_labels (labels: list label) : bool =
  not (has_duplicates labels [])
\end{lstlisting}
 
The \texttt{seen} accumulator corresponds precisely to $\Sigma$; the \texttt{RECUR} label is excluded because it is a reserved transition marker that may appear on multiple edges by design.
Z3 discharges the bridge lemma automatically, since the implication reduces to a statement about list membership over a finite accumulator.
 
This is precisely the point at which \textsc{Platum} and DATUM diverge most sharply.
In DATUM, label uniqueness is enforced \emph{intrinsically}: the \texttt{Choice} constructor requires the user to pass the accumulated label history as an explicit argument, and F* discharges a \texttt{\{no\_duplicate\_labels l\}} refinement obligation on every branch addition.
The consequence is that the protocol definition is structurally coupled to its own well-formedness proof: every edit to the session type re-triggers an SMT check over the entire label history, and the \texttt{Choice} constructor exposes arguments that carry no semantic meaning in the protocol itself.
In \textsc{Platum}, the session type is written with only the five semantic inputs; \texttt{check\_unique\_labels} is invoked once, after the fact, as a standalone pass over the finished AST.
The protocol engineer sees no accumulators, no proof scaffolding, and no coupling between specification edits and proof re-checking.
 
\bigskip
\noindent\textbf{Check 2: Guarded Recursion.}\quad
The procedure $\mathcal{A}_{guarded}$ confirms that every path from a $\mu$-binder to a corresponding \texttt{recur} includes at least one communication step, preventing infinite internal divergence.
Letting $\xrightarrow{\epsilon^*}$ denote a sequence of internal (non-communicating) transitions, the check confirms there is no path from a recursion variable $X$ back to itself consisting only of such steps:
$$
\nexists\, \text{path}:\ X \xrightarrow{\epsilon^*} X
$$
Intuitively, this prevents the synthesized monitor from spinning indefinitely without observing any message, which would make the FSM unresponsive to network traffic.
Because continuations are inspectable F* functions, the procedure evaluates each $\mu$-binder body by applying it to a fresh variable, making the check compositional across nested binders.
 
\bigskip
\noindent\textbf{Check 3: Global Progress.}\quad
The procedure $\mathcal{A}_{progress}$ confirms that the protocol never reaches a stuck state.
Every non-terminal state in the intermediate representation graph $P$ must possess at least one valid outgoing transition:
$$
\forall s \in P_{states}.\ s_{isEnd} \vee |s_{transitions}| > 0
$$
 
\bigskip
\noindent\textbf{Check 4: Session Fidelity.}\quad
The procedure $\mathcal{A}_{valid}$ confirms that the protocol is closed with respect to state transitions: every transition target must exist within the defined graph.
$$
\forall s \in P_{states},\ \forall t \in s_{transitions}.\ t_{target} \in \{ s'.id \mid s' \in P_{states} \}
$$
This check ensures the synthesized C monitor will never jump to an undefined label or execute an illegal transition.
 
\medskip
Together, these four checks confirm that any session type accepted by the \textsc{Platum} well-formedness suite is structurally ready for synthesis: no ambiguous branch points, no unproductive loops, no stuck states, and no dangling state references.
Because all four procedures run before C synthesis begins, the generated FSM inherits these invariants by construction, and the synthesis step operates on a graph already known to satisfy the structural prerequisites for correct centralized enforcement.
%%% Local Variables:
%%% mode: LaTeX
%%% TeX-master: t
%%% End:
\section{Synthesis of Performant Protocol Monitors}
\label{sec:synthesis}

\subsection{Extraction Semantics}
\label{subsec:extraction}

Synthesis in \textsc{Platum} operates on session types that have already passed the compile-time well-formedness checks described in Section~\ref{sec:methodology}.
The input to the synthesis pipeline is a structurally confirmed global RMPST whose AST is available for inspection via Meta-F*.
The structural guarantees established by the decision procedures (label uniqueness, guarded recursion, progress, and fidelity) are preconditions that the synthesis process inherits: no additional structural reasoning is required during translation.

A central challenge in our synthesis pipeline is performing type erasure on the F* terms without discarding the vital logical constraints embedded within them.
F* is a dependently typed language where a value's type can depend on runtime data (e.g., $n:\text{int}\{n > 3\}$), whereas C is simply typed.
To bridge this gap, we employ Meta-F*'s reflection capabilities.
We inspect the refinement terms inside the DSL AST and confirm that they are decidable boolean propositions.
We then extract these terms and transform them into C syntax trees.
Effectively, a dependent type in F* is transformed into a standard C variable paired with a control-flow guard.
The refinement $n > 3$ is stripped from the type signature but immediately re-inserted as an \texttt{if(n > 3)} check in the generated state machine.

The extraction process begins by analyzing the functional binders of the protocol specification. In a dependently typed setting, a binder $b$ carries a sort $\tau$ that may contain logical predicates constraining the value space. To synthesize a C monitor, we must decouple the storage requirement (the base type) from the validation requirement (the refinement logic). We define an extraction function $\mathcal{E}(b)$ that operates on the term view of the binder's sort using Meta-F* reflection. This function inspects the structure of $\tau$ to detect the presence of the \texttt{Tv\_Refine} constructor. If a refinement is found, we isolate the base type $\tau_{base}$, which maps to standard C types like \texttt{int32\_t}, and the refinement formula $\phi$, which represents the logical constraint.

\[
\mathcal{E}(b) = \begin{cases} 
(x, \tau_{base}, \mathcal{R}(\phi)) & \text{if } \text{inspect}(b.\text{sort}) = \text{Tv\_Refine}(x: \tau_{base}, \phi) \\
(x, \tau, \text{true}) & \text{otherwise}
\end{cases}
\]

When a refinement $\phi$ is detected, the reification function $\mathcal{R}$ is invoked to traverse the term. 
By mapping F* primitive operators (e.g., \texttt{Prims.op\_GreaterThan}) to their corresponding constructors in our intermediate representation, we ensure that the logical constraints confirmed by the decision procedures are preserved exactly in the AST used for synthesis. This guarantees that the generated runtime checks are not merely heuristics, but faithful implementations of the structurally confirmed specification.

The output of the extraction pass is an intermediate representation(IR) graph that serves two purposes. First, it materializes the abstract syntax tree used by Meta-F* into a concrete, inspectable structure, which is a necessary step because the IO operations that write the final \texttt{monitor.c} file rely on an OCaml runtime which cannot directly reason over F* tactic terms; the graph provides the concrete artifact they operate on.
Second, it makes the structural correspondence between the session type and the FSM explicit and checkable: session type choices become transitions, session states become nodes, and the well-formedness properties confirmed in Section~\ref{subsec:wellformedness} apply directly to this graph before a single line of C is generated.
The jump from a dependently typed F* specification to a flat C state machine is otherwise large enough to obscure whether the structural invariants survive translation. This IR graph  makes that correspondence explicit, and the direct translation in Section~\ref{subsec:translation} then operates on a representation whose properties are already known.
The graph carries no proof obligations of its own and is not exposed to the user; its theoretical properties as a formal object are the subject of ongoing work.

\subsection{Direct Translation Semantics}
\label{subsec:translation}

Once the refinement logic is isolated, we must bridge the semantic gap between the constructive logic of F* and the strict boolean evaluation of C.  In the generated C monitor a variable $v$ must be resolved to a specific memory location: either a transient field in the current message stack frame (e.g., \texttt{payload.v}) or a persistent state variable stored in the monitor's context heap (e.g., \texttt{mon->ctx.v}).
 For a given refinement term $\phi$, the translation $\mathcal{T}(\phi, \Gamma)$ operates within a variable context $\Gamma$, distinguishing between variables local to the current message payload ($\Gamma_{local}$) and those persisting in the monitor state. The translation rules for boolean connectives and arithmetic operators are defined as follows, where $\parallel$ denotes string concatenation:

\[
\mathcal{T}(t, \Gamma) = \begin{cases} 
    \text{"("} \parallel \mathcal{T}(l, \Gamma) \parallel \text{" \&\& "} \parallel \mathcal{T}(r, \Gamma) \parallel \text{")"} & \text{if } t = \texttt{op\_logical\_and}\ l\ r \\
    \text{"("} \parallel \mathcal{T}(l, \Gamma) \parallel \text{" || "} \parallel \mathcal{T}(r, \Gamma) \parallel \text{")"} & \text{if } t = \texttt{op\_logical\_or}\ l\ r \\
    \mathcal{T}(l, \Gamma) \parallel \text{" == "} \parallel \mathcal{T}(r, \Gamma) & \text{if } t = \texttt{op\_Equality}\ l\ r \\
    \text{"!"} \parallel \mathcal{T}(e, \Gamma) & \text{if } t = \texttt{op\_Negation}\ e \\
    \texttt{payload.}v & \text{if } t = \texttt{Tv\_Var}\ v \land v \in \Gamma_{local} \\
    \texttt{mon->ctx.}v & \text{if } t = \texttt{Tv\_Var}\ v \land v \notin \Gamma_{local}
\end{cases}
\]

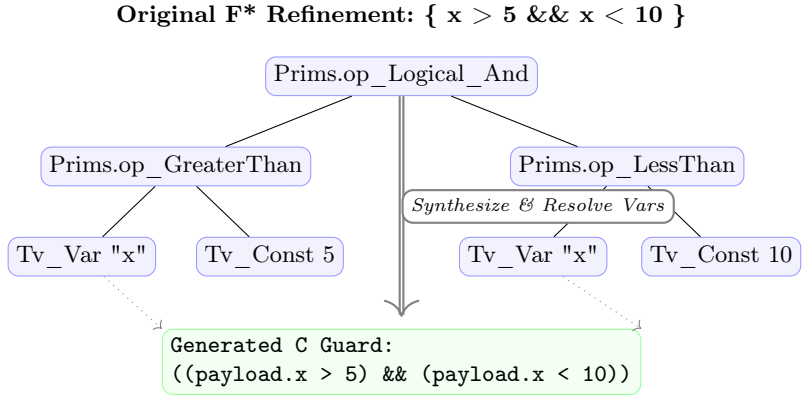
\begin{figure}[h]
    \centering
    \begin{tikzpicture}[
        level 1/.style={sibling distance=6cm}, % WIDER sibling distance
        level 2/.style={sibling distance=2.5cm},
        level distance=1.2cm,
        every node/.style={draw, rectangle, rounded corners, align=center, font=\footnotesize, fill=white},
        astnode/.style={fill=blue!5, draw=blue!40},
        codeblock/.style={draw=green!40, fill=green!5, font=\ttfamily\footnotesize, align=left}
    ]

    % Header Refinement Text (Moved closer to root)
    \node[draw=none, fill=none, font=\bfseries] at (0, 0.8) {Original F* Refinement: \{ x > 5 \&\& x < 10 \}};

    % Root of the AST (Meta-F*)
    \node[astnode] (root) {Prims.op\_Logical\_And}
        child {
            node[astnode] (left) {Prims.op\_GreaterThan}
            child { node[astnode] {Tv\_Var "x"} }
            child { node[astnode] {Tv\_Const 5} }
        }
        child {
            node[astnode] (right) {Prims.op\_LessThan}
            child { node[astnode] {Tv\_Var "x"} }
            child { node[astnode] {Tv\_Const 10} }
        };

    % Arrow indicating transformation (Shortened distance)
    \draw[->, double, thick, gray] (root) -- ++(0,-3.2) node[midway, right, text=black, font=\scriptsize] {\textit{Synthesize \& Resolve Vars}};

    % The resulting C Code (Moved closer to AST)
    \node[codeblock, below of=root, node distance=3.8cm, minimum width=6cm] (c_code) {
        \textbf{Generated C Guard:}\\
        \texttt{((payload.x > 5) \&\& (payload.x < 10))}
    };

    % Connecting lines directly from leaves to C code conceptual parts
    \draw[dotted, gray, ->] (left-1) -- (c_code.north west);
    \draw[dotted, gray, ->] (right-1) -- (c_code.north east);

    \end{tikzpicture}
    \caption{The reification pipeline. The meta-program inspects the F* AST (top), mapping logical operators like \texttt{op\_Logical\_And} to C's \texttt{\&\&}, and resolving the variable \texttt{x} to its runtime location \texttt{payload.x}.}
    \label{fig:reification}
\end{figure}

This context-sensitivity is the cornerstone of our synthesis safety argument. By explicitly resolving variable locations during translation, we prevent the "scope confusion" errors common in manual parser implementation. Figure \ref{fig:reification} visualizes this transformation pipeline, tracing the path from the raw F* AST nodes for the refinement \texttt{\{x > 5 \&\& x < 10\}} down to the final executable C conditional. Note how the abstract variable \texttt{x} is concretized into a structure member access \texttt{payload.x} in the final output.

We define the monitor state as a C structure containing the context variables identified during the AST analysis.
The logic is encapsulated in a stateless \texttt{step} function that acts as a transition function for the automata.
This function takes the current monitor state and an incoming MAVLink message, decodes the payload, and evaluates the guard conditions derived from the refinements.
If the guard holds (e.g., the refinement returns true), the monitor updates its internal context variables (simulating the passing of arguments to the recursive function in F*) and transitions to the next state ID.
If the guard fails, the monitor flags a violation.
This direct mapping ensures that the synthesized code is a faithful execution of the formal model.

\begin{figure}[t]
    \centering
\begin{lstlisting}[language=C, caption={Synthesized C Monitor Logic for the Mission Count state. Note how the F* refinement is translated into a C conditional check, and the recursive argument passing is compiled into a context update.}, label={lst:c_monitor}]
bool monitor_step(monitor_t* mon, const mavlink_message_t* msg) {
  switch (mon->state) {
// State 0: Expecting MISSION_COUNT
// GCS -> UAV: MISSION_COUNT(N: Z {N >= 1 /\ N < LIMIT})
    case STATE_IDLE_0: {
      if (msg->msgid == MAVLINK_MSG_ID_MISSION_COUNT) {
        mavlink_mission_count_t payload;
        mavlink_msg_mission_count_decode(msg, &payload);
// Check Refinement: {cnt >= 1 && cnt < mission_item_limit}
        if(payload.count >= 1 && payload.count < 65535) {
           // Update Context (F* bound variables)
           mon->ctx.cnt = payload.count;
           mon->ctx.curr = 0; // Initialize Mu variable
// Transition
           mon->state = STATE_LOOP_REQ_1;
           return true;
        }
      }
      break; 
      // ...
  }
}
\end{lstlisting}
\end{figure}

\subsection{Scalability and Correctness Rationale}

\textsc{Platum} targets a flat Finite State Machine (FSM) architecture over DATUM's OCaml extraction.
This choice eliminates the managed runtime entirely: no garbage collector, no runtime type information, no non-deterministic pauses.
In a hard real-time system, GC pauses of even a few milliseconds can violate timing deadlines; by synthesizing raw C, \textsc{Platum} relies solely on the OS scheduler, yielding the deterministic performance profile presented in Table~\ref{tab:results}.
The flat FSM architecture also guarantees a bounded Worst-Case Execution Time and $O(1)$ memory usage relative to mission duration, with no dynamic allocation or deep recursion in the generated code.
This is directly suited to \textsc{Platum}'s centralized enforcement model: a single FSM instance runs at the GCS/UAV network boundary, and the total enforcement overhead is concentrated at one point and can be precisely bounded, as Section~\ref{sec:evaluation} confirms.
To further minimize impact on high-frequency telemetry, \textsc{Platum} implements an optimized \emph{Pre-Filter} stage.
MAVLink streams are dominated by high-rate messages irrelevant to mission logic; an $O(1)$ look-up table whitelists non-critical messages before the full state machine is consulted.

A natural question is what correctness guarantee the synthesized monitor provides end-to-end.
Informally, the synthesis pipeline preserves the semantics of the confirmed session type: each C state corresponds to a node in the IR graph, each guard condition is a faithful C translation of the corresponding F* refinement predicate (established by the extraction semantics in Section~\ref{subsec:extraction}), and each state transition corresponds to an edge that passed the fidelity check.
Any message sequence accepted by the monitor therefore constitutes a valid trace of the global session type.
A formal proof of this monitor soundness property, connecting the boolean output of the C \texttt{step} function to trace inclusion in the global type's semantics, is deferred to future work.
The present pipeline provides the structural preconditions for such a result: a monitor synthesized from a well-formed global RMPST cannot transition to an undefined state, cannot accept a payload that violates its refinement, and cannot reach a configuration from which no further progress is possible.

One might ask why we do not employ a verified extraction pipeline like Low*, which is standard in the F* ecosystem for generating C.
Low* is most valuable when the target code manipulates heap-allocated buffers and requires pointer aliasing or ownership proofs, as in cryptographic libraries such as EverCrypt.
Our synthesized monitors present none of these concerns: \texttt{monitor\_ctx\_t} is a fixed-size value type, \texttt{monitor\_t} is caller-allocated and passed by pointer, and there is no dynamic dispatch or heap involvement in the generated code.
The C our pipeline produces would be structurally identical to what KreMLin would extract from an equivalent Low* program, at a fraction of the specification cost.
We therefore bypass Low* in favor of a direct AST-to-C translation, retaining the logical correctness of the protocol confirmed via the RMPST well-formedness checks, without burdening the synthesis pipeline with heap reasoning that does not apply to our static architecture.

%%% Local Variables:
%%% mode: LaTeX
%%% TeX-master: t
%%% End:
\section{Runtime Monitoring and Overhead Analysis}
\label{sec:evaluation}

We empirically validated the performance improvements of \textsc{Platum} through a series of comparative experiments using the ArduPilot Software-In-The-Loop (SITL) simulation.
Our evaluation focuses on quantifying the overhead introduced by the monitor in terms of Resident Set Size (RSS) memory consumption and latency, the two critical performance metrics for resource-constrained avionics middleware.
All experiments were conducted on a Mac mini equipped with an Apple M4 Chip and 16 GB of unified memory, running macOS Sequoia 15.7.3, with ArduCopter V4.7.0-dev in SITL mode communicating with QGroundControl (v5.0.8).
While the experiments were performed in a simulated environment, the synthesized C 
monitor is platform-agnostic and links directly into any native MAVLink communication 
stack, including ARM-based flight controller firmware (e.g., STM32 running PX4 or 
ArduPilot). The Python proxy used here is an evaluation harness, not an architectural 
requirement: a production deployment would replace it with a native C or C++ MAVLink 
proxy, at which point the \texttt{ctypes} FFI overhead is eliminated entirely and 
System Latency converges to the Monitor Latency figures reported in 
Table~\ref{tab:results}.
\subsection{Experimental Setup}

The experimental architecture consists of a custom Python-based MAVLink proxy interposed between the ArduPilot SITL instance and the GCS, evaluated under two configurations:

\begin{enumerate}
    \item \textbf{Baseline Configuration:} The proxy operates in ``passthrough'' mode, parsing MAVLink packets using the standard Python bindings but performing no enforcement logic before forwarding.
    This establishes a control baseline for latency inherent to the network stack and OS scheduling.
    \item \textbf{Platum Configuration:} The proxy utilizes our synthesized C monitor, compiled as a shared dynamic library (\texttt{.dylib}).
    Every incoming packet from the GCS is passed to the C monitor's step function via \texttt{ctypes} bindings, and forwarded to the UAV only if the monitor returns \texttt{True}.
\end{enumerate}

We simulated a mission upload sequence of N=100 repeated waypoint transactions.
This scenario exercises the recursive logic of the monitor and generates sustained traffic, stress-testing the monitor's state transition throughput.
Memory usage was measured via the OS \texttt{getrusage} syscall to track Peak RSS, while latency was captured using monotonic clocks wrapping the C library call.

\subsection{Comparative Results}

The results, summarized in \autoref{tab:results}, demonstrate a substantial improvement in resource efficiency.
We categorize latency into two metrics: \textit{Monitor Latency} (pure FSM step cost) and \textit{System Latency} (total overhead added to the proxy pipeline).

\textbf{Latency Analysis:} \textsc{Platum} introduces a negligible \textit{Monitor Latency} of approximately 0.12 $\mu s$ (123 ns) in its idle state and 3.88 $\mu s$ during active monitoring, confirming that state machine transitions are effectively instantaneous relative to network speed.
Accounting for full system overhead, the total added latency is 13.33 $\mu s$, compared to 56.74 $\mu s$ for DATUM, yielding an approximately $4\times$ reduction.
The gap between Monitor and System Latency (approximately 9 $\mu s$) is attributable entirely to the \texttt{ctypes} FFI bridging and Python interpreter overhead, not to the centralized monitor itself.

\textbf{Memory Efficiency:} The RSS Delta of the \textsc{Platum} setup is approximately 8.39 MB, the majority of which is administrative overhead from the Python FFI and buffer copies rather than the C monitor itself.
The C monitor's static context structure (\texttt{monitor\_ctx\_t}) requires only a few hundred bytes.
DATUM reported an RSS increase of over 13 MB due to the OCaml runtime.
Because \textsc{Platum} does not distribute monitoring across participants, the reported overhead is the complete cost of centralized enforcement at the GCS/UAV boundary, not a per-participant contribution.

\begin{table}[t]
    \centering
    \caption{Performance comparison: \textsc{Platum} (Optimized C) vs. DATUM (OCaml). \textsc{Platum} achieves a significant reduction in system latency and lower memory pressure.}
    \label{tab:results}
    \begin{tabular}{|l|c|c|c|c|}
        \hline
        \textbf{System} & \textbf{Language} & \textbf{RSS Delta} & \textbf{Total Latency} & \textbf{Logic Delta} \\
        \hline
        DATUM \cite{amorimEnforcingMAVLinkSafety2025} & OCaml & $\sim$13.72 MB & 56.74 $\mu s$ & 27.14 $\mu s$ \\
        \hline
        Proxy Baseline & Python & 8.39 MB & 9.11 $\mu s$ & N/A \\
        \textbf{Platum (Ours)} & \textbf{C} & \textbf{8.39 MB} & \textbf{13.33 $\mu s$} & \textbf{4.22 $\mu s$} \\
        \hline
    \end{tabular}
\end{table}

\subsection{Performance Impact and Architectural Advantages}

The performance gap between \textsc{Platum} and DATUM stems primarily from the elimination of the OCaml runtime.
OCaml requires a garbage collector and a complex runtime heap, introducing non-deterministic pauses that are antithetical to real-time UAV control loops.
By synthesizing monitors into pure, stack-allocated C code, every state transition has a constant, predictable execution time.
The monitor operates directly on decoded MAVLink structures, bypassing the expensive object-mapping found in OCaml-to-Python bridges.

The difference in system latency between the proxy baseline and the active monitor is only 4.22 $\mu s$, while DATUM's RSS delta relative to its baseline exceeded 13,728 KB.
Our approach keeps memory overhead substantially lower while providing stronger formal guarantees through the $F^*$ synthesis pipeline.

The current evaluation covers the MAVLink Mission Upload sub-protocol, which exercises recursive state transitions, sequenced refinement checks, and sustained traffic: the conditions most relevant to stress-testing the centralized monitor's decision throughput.
Extending \textsc{Platum} to the broader MAVLink microservice suite requires authoring new global session type specifications in the DSL; the synthesis pipeline, the C FSM runtime, and the proxy deployment infrastructure are fully reusable across protocols.
Coverage of additional services is a natural direction for future work.
\section{Related Work}

\paragraph{Physics-Based Anomaly Detection (R2U2)}
Systems like R2U2~\cite{R2U2MonitoringDiagnosis} utilize models of the UAV's physical dynamics to detect intrusions.
While effective against kinematic attacks, these methods cannot detect logic attacks where physical behavior remains nominal (e.g., logic bombs in mission uploads).
\textsc{Platum} specifically targets these logical vulnerabilities that physics-based models miss~\cite{khanUnsupervisedAnomalyDetection2019}.

\paragraph{Global Session Type Tooling (Scribble, NuSCR)}
Scribble~\cite{yoshidaScribbleProtocolLanguage2014} and the NuSCR toolchain~\cite{zhouRefiningMultipartySession} support authoring global multiparty session type specifications and generating endpoint code via projection to local types.
NuSCR specifically targets F* and Go, supporting interaction refinements with SMT-backed predicates on message payloads.
\textsc{Platum} shares the global specification model but differs in two key respects.
First, the \textsc{Platum} DSL is embedded in F* with Meta-F* reflective AST inspection, enabling algorithmic well-formedness checking without the user-facing proof terms required by intrinsic approaches.
Second, \textsc{Platum} does not project to local types: the global session type is compiled directly to a centralized C FSM, which is the appropriate enforcement architecture for UAV proxy monitoring where participant-side instrumentation is not feasible.

\paragraph{Runtime Monitoring Languages (RML, Varanus)}
The Runtime Monitoring Language~\cite{anconaRMLTheoryPractice2021} is a domain-specific language for synthesizing monitors from specifications over structured event sequences.
RML monitors are expressive enough to handle non-context-free properties and have been applied to robotics environments via ROSMonitoring.
Similarly, Varanus~\cite{luckcuckVaranusRuntimeVerification2025} uses Communicating Sequential Processes (CSP) to synthesize oracles that monitor ordered event sequences in autonomous rovers.
The key distinctions are that neither framework carries refinement predicates tied to message payload values in the same session-typed manner as \textsc{Platum}, and their generated monitors rely on higher-level runtimes rather than bare-metal embedded C.
\textsc{Platum}'s use of global RMPSTs provides the data-dependent payload constraints essential for MAVLink well-formedness while targeting resource-constrained deployment.

\paragraph{Symbolic Protocol Analysis (Sprout)}
The Sprout verifier~\cite{liSproutVerifierSymbolic2025} represents concurrent and independent work in symbolic protocol implementability.
Sprout is a network-parametric tool that checks Symbolic Coherence Conditions to determine whether a global protocol, including those with dependent refinements on message values, can be soundly realized as a distributed system across different network architectures (FIFO, Bag, Mailbox).
While Sprout addresses the theoretical implementability question, \textsc{Platum} provides a framework focusing on the practical synthesis of centralized proxy monitors for embedded C deployment.

\paragraph{Cryptographic and Static Security (MAVSec, PAVE)}
Early UAV security research focused on confidentiality through encryption (MAVSec~\cite{allouchMAVSecSecuringMAVLink2019}, MAV-DTLS~\cite{chaariMAVDTLSSecurityEnhancement2020}) or static design-level analysis (PAVE~\cite{wrayPAVEMAVLinkFormalVerification2025}, Tamarin~\cite{meierTAMARINProverSymbolic2013}).
While powerful, these approaches treat the message payload as a black box or are limited to offline analysis.
\textsc{Platum} complements them by bringing live, logic-aware centralized monitors to the GCS/UAV communication boundary, enforcing value-dependent ordering invariants that static and cryptographic tools do not address.

\paragraph{Prior Session Type-based MAVLink Monitoring (DATUM)}
Our work builds on DATUM~\cite{amorimEnforcingMAVLinkSafety2025,taylorSecuringModbusBasedIndustrial2025}, which first used global RMPSTs as a specification language for centralized MAVLink proxy enforcement.
However, DATUM's intrinsic verification strategy required users to supply accumulated label histories, recursion context stacks, and proof scaffolding as part of the session type itself, obscuring the five semantic components beneath a significant burden of proof terms.
Additionally, its OCaml extraction backend incurred high RSS overhead and introduced non-deterministic GC pauses incompatible with hard real-time UAV control loops.
\textsc{Platum} addresses both limitations: extrinsic well-formedness checking via reflective decision procedures in Meta-F* eliminates the proof burden, and direct C FSM synthesis eliminates the managed runtime entirely.

\paragraph{Verified C Extraction (Low*)}
The Low*~\cite{protzenkoVerifiedLowlevelProgramming2017} language is the standard for extracting verified C from F*, enabling memory-safety proofs for complex heap-manipulating programs such as cryptographic libraries.
We deliberately bypass Low* because our target domain (flat, allocation-free FSMs) comprises a strict subset of C with no dynamic memory allocation, no pointer aliasing, and no heap involvement; the direct AST-to-C translation in \textsc{Platum} produces structurally identical code at a fraction of the specification cost.

\section{Conclusion}

As UAVs proliferate in safety-critical infrastructure, the security of their communication protocols demands enforcement mechanisms that are both formally grounded and practically deployable.
Prior work (DATUM) established global RMPSTs as an effective specification language for centralized MAVLink proxy enforcement, but its intrinsic verification strategy imposed a prohibitive specification burden, and its OCaml extraction backend introduced RSS overhead and non-deterministic GC pauses that are incompatible with hard real-time UAV control loops.

By separating specification from well-formedness checking, using reflective decision procedures in Meta-F* to confirm structural invariants over the protocol AST rather than requiring manual proof terms during term construction, \textsc{Platum} reduces the specification barrier to its minimum: five inputs per interaction step.
By targeting direct C FSM synthesis rather than OCaml extraction, the framework produces centralized proxy monitors that eliminate managed runtime overhead and are compatible with ARM microcontroller deployment.
\textsc{Platum} demonstrates that global RMPSTs, used as a specification language for centralized runtime enforcement, can deliver both formal structural guarantees and the real-time performance demanded by safety-critical UAV applications.
The path toward a full metatheoretic treatment and decentralized runtime checking, covering projection, local type correctness, and end-to-end trace soundness, remains open as a natural continuation of this work.

%%% Local Variables:
%%% mode: LaTeX
%%% TeX-master: t
%%% End:

\bibliographystyle{splncs04}
\bibliography{bib-4}

\end{document}